\documentclass[10pt]{iopart}

\usepackage{latexsym}
\usepackage{amssymb}
\usepackage{cite}

\newcommand{\be}{\begin{equation}}
\newcommand{\ee}{\end{equation}}

\newcommand{\beqn}{\begin{eqnarray}}
\newcommand{\eeqn}{\end{eqnarray}}
\newcommand{\bea}{\begin{eqnarray}}
\newcommand{\eea}{\end{eqnarray}}

\newcommand{\ba}{\begin{array}}
\newcommand{\ea}{\end{array}}

\def \BE {\begin{equation}}
\def \EE {\end{equation}}

\def \bl {\mbox{\boldmath{$\ell$}}}
\def \hbl {\mbox{\boldmath{$\hat \ell$}}}
\def \bn {\mbox{\boldmath{$n$}}}
\def \bm #1 {\mbox{\boldmath{$m^{({#1})}$}}}
\def \bmd #1 {\mbox{\boldmath{$m_{({#1})}$}}}
\def \hbn {\mbox{\boldmath{$\hat n$}}}
\def \hbm #1 {\mbox{\boldmath{$\hat m^{({#1})}$}}}

\def \hL {\hat{L}}
\def \hN {\hat{N}}

\newcommand{\M}[3] {{\stackrel{#1}{M}}_{{#2}{#3}}}
\newcommand{\hM}[3] {{\stackrel{#1}{\hat{M}}}_{{#2}{#3}}}

\def \hbm #1 {\mbox{\boldmath{$\hat m^{(#1)}$}}}

\def \BEAH {\begin{eqnarray*}}
\def \EEAH {\end{eqnarray*}}
\def \BEA {\begin{eqnarray}}
\def \EEA {\end{eqnarray}}
\def \BDM {\begin{displaymath}}
\def \EDM {\end{displaymath}}
\def \Mi {\stackrel{i}{M}}
\def \Mj {\stackrel{j}{M}}

\def \pul {{\textstyle{\frac{1}{2}}}}

\def \ms {\mbox{\boldmath$\sigma$}}

\def \T {\bigtriangleup  }

\def \S {S}

\usepackage{latexsym}
\newcommand{\qed}{
\ifmmode \quad\Box
\else
\leavevmode\unskip\penalty9999 \hbox{}\nobreak\hfill
\quad\hbox{$\Box$}
\fi}

\newtheorem{proposition}{Proposition}

\def \eq5d {\stackrel{{\mbox{\tiny 5D}}}{=}}
\def \kd #1 {\delta_{#1}}
\def \beah {\begin{eqnarray*}}
\def \eeah {\end{eqnarray*}}

\def \WDS #1 {\mbox{$\Phi_{#1}^{S}$}}
\def \WDA #1 {\mbox{$\Phi_{#1}^{A}$}}
\def \WD #1 {\mbox{$\Phi_{#1}$}}

\begin{document}
\title{Ricci identities in higher dimensions}

\author{M. Ortaggio$^1$, V. Pravda$^2$ and A. Pravdov\' a$^2$}

\address{$^1$ Dipartimento di Fisica, Universit\`a degli Studi di Trento, 
	and INFN, \\ Gruppo Collegato di Trento, Via Sommarive 14, 38050 Povo (Trento), Italy  }
\address{$^2$ Mathematical Institute, 
Academy of Sciences, \v Zitn\' a 25, 115 67 Prague 1, Czech Republic}
\eads{\mailto{marcello.ortaggio `AT' seznam.cz}, \mailto{pravda@math.cas.cz}, \mailto{pravdova@math.cas.cz}}

\begin{abstract}
We explore connections between geometrical properties of null congruences and the algebraic structure of the Weyl tensor in $n>4$ spacetime dimensions. First, we present the full set of Ricci identities on a suitable ``null'' frame, thus completing  the extension of the Newman-Penrose formalism to higher dimensions. Then we specialize to geodetic null congruences and study specific consequences of the Sachs equations. These imply, for example, that Kundt spacetimes are of type II or more special (like for ${n=4}$) and that for odd $n$ a twisting geodetic WAND must also be shearing (in contrast to the case ${n=4}$).
\end{abstract}

\pacs{04.50.+h, 04.20.-q, 04.20.Cv}
% 		Gravity in more than four dimensions; Classical general relativity; Fundamental problems and general formalism

\vspace{5mm}
\noindent
{\small{ {\bf{Changes with respect to the published version:}} typo in the last term in the first line of (11f) corrected,  missing term on the r.h.s. of (11p) added, first sentence between Propositions 2 and 3 slightly changed.} }
\section{Introduction}
\label{sec_intr}

Thanks to the correspondence between geometrical properties of null geodesics and optical properties of the gravitational field, the study of ray optics (see, e.g., \cite{Stephanibook,penrosebook2} and references therein) has played a major role in the construction and classification of exact solutions of Einstein's equations in $n=4$ dimensions. In this context, a fundamental connection between geometric optics and the algebraic structure of the Weyl tensor is provided by the Goldberg-Sachs theorem \cite{GolSac62,NP}, which states that a vacuum metric is algebraically special if and only if it contains a shearfree geodetic null congruence (cf.~\cite{Stephanibook,penrosebook2} for related results and generalizations). 

In recent years, possible extensions of the above concepts to arbitrary dimensions $n>4$ have been investigated. The classification of the Weyl tensor has been presented in \cite{Coleyetal04,Milsonetal05}, and aspects of geometric optics have been studied in \cite{FroSto03,Pravdaetal04,LewPaw05,PodOrt06}. In particular, a partial extension of the Goldberg-Sachs theorem to $n>4$ has been proven in \cite{Pravdaetal04} by considering various contractions of the Bianchi identities. One of the results was that the multiple\footnote{``Multiple'' WANDs are those WANDs whose order of alignment \cite{Coleyetal04,Milsonetal05} with the Weyl tensor is greater than zero.} WAND (Weyl aligned null direction) of a type III or N vacuum spacetime is necessarily geodetic. Nevertheless, possible ``violations'' of the Goldberg-Sachs theorem in $n>4$ have been also pointed out. For example, the principal null directions of $n=5$ rotating black holes (which are of type D) are geodetic but shearing \cite{FroSto03,Pravdaetal04}. Furthermore, in $n>4$ vacuum spacetimes of type III or N, a multiple principal null direction with expansion necessarily has also non-zero shear \cite{Pravdaetal04}. 

This contribution makes further progress in the study of optical properties of the gravitational field in higher dimensions. After summarizing our notation and presenting a few preliminary results (section~\ref{sec_preliminary}), we derive the full set of frame Ricci identities in $n\ge4$ (section~\ref{sec_ricci}). This is a natural complement to the frame Bianchi identities given in \cite{Pravdaetal04}, and together with the expressions for commutators given in section~1.2 of \cite{Coleyetal04vsi} it generalizes the Newman-Penrose formalism \cite{NP,Stephanibook,penrosebook2} for any $n>4$. 
Consequently, one can explore which of the four-dimensional applications of this formalism can be extended to higher dimensions.
Here, we focus on some of the consequences of the Ricci identities, such as an analog of the Sachs equations \cite{Sachs61,Stephanibook,penrosebook2} for geodetic null congruences and their implications (section~\ref{sec_sachs}).

\section{Null frames and Ricci rotation coefficients}

\label{sec_preliminary}

In an $n$-dimensional spacetime we can set up a frame of $n$ real vectors $\bm{a} $ which consists of two null vectors $\bm{0} =\bn$, $\bm{1} =\bl$ and $n-2$ orthonormal spacelike vectors $\bm{i} $  \cite{Pravdaetal04}. From now on, $a, b\dots=0,\dots,n-1$ while $i, j  \dots=2,\ldots,n-1$. We will observe Einstein's summation convention for both types of indices. For indices $i, j, \dots$ there is no difference between covariant and contravariant components and thus we will not distinguish between subscripts and superscripts.
The metric reads $g_{a b} = 2\ell_{(a}n_{b)} + \delta_{ij} m^{(i)}_a m^{(j)}_b$ so that the scalar products of the basis vectors are 
\be
	\hspace{-1.5cm} \ell^a \ell_a= n^a n_a =\ell^a m^{(i)}_{a}=n^a m^{(i)}_a= 0, \qquad  \ell^a n_a = 1, \qquad
 		m^{(i)a}m^{(j)}_a=\delta_{ij}. 
 	\label{products}	
\ee 

\subsection{Ricci rotation coefficients}

We are interested in properties of the covariant derivatives of the above frame vectors. 
Let us define the Ricci rotation coefficients $L_{ab}$, $N_{ab}$ and $\Mi_{ab}$ by
\be
\fl 
\ell_{a;b}=L_{cd} m^{(c)}_a m^{(d)}_{b} \ , \qquad n_{a;b}=N_{cd} m^{(c)}_a m^{(d)}_{b} \ , \qquad
m^{(i)}_{a;b}=\Mi_{cd} m^{(c)}_a  m^{(d)}_{b}  \ .\label{derlnm}
\ee

First derivatives of eqs.~(\ref{products}) lead to the following $n^2(n+1)/2$ relations \cite{Pravdaetal04}
\be 
\fl
 L_{0a}=N_{1a}=N_{0a}+L_{1a}=\Mi_{0a} + L_{ia} = \Mi_{1a}+N_{ia}=\Mi_{ja}+\Mj_{ia}=0 , 
 \label{const-scalar-prod}
\ee
which reduce to $n^2(n-1)/2$ the number of independent rotation coefficients\footnote{In $n=4$ dimensions the $L_{ab}$, $N_{ab}$ and $\Mi_{ab}$ are equivalent to the 12 complex Newman-Penrose spin coefficients. Namely, with the notation $\sqrt{2}\mbox{\boldmath{$m$}}=\bm{2} -i\bm{3} $ one has (different conventions in the literature may affect certain coefficient signs): $-\sqrt{2}\kappa\equiv \sqrt{2}\ell_{a;b}m^a\ell^b=L_{20}-iL_{30}$, $-2\rho\equiv 2\ell_{a;b}m^a\bar{m}^b=L_{22}+L_{33}+2iL_{[23]}$, $-2\sigma\equiv 2\ell_{a;b}m^a m^b=L_{22}-L_{33}-2iL_{(23)}$, $-\sqrt{2}\tau\equiv \sqrt{2}\ell_{a;b}m^a n^b=L_{21}-iL_{31}$, $\sqrt{2}\nu\equiv \sqrt{2}n_{a;b}\bar{m}^a n^b=N_{21}+iN_{31}$, $2\mu\equiv 2n_{a;b}\bar{m}^am^b=N_{22}+N_{33}-2iN_{[23]}$, $2\lambda\equiv 2n_{a;b}\bar{m}^a\bar{m}^b=N_{22}-N_{33}+2iN_{(23)}$, $\sqrt{2}\pi\equiv \sqrt{2}n_{a;b}\bar{m}^a\ell^b=N_{20}+iN_{30}$, $\varepsilon+\bar\varepsilon\equiv -\ell_{a;b}n^a\ell^b=-L_{10}$, $\varepsilon-\bar\varepsilon\equiv m_{a;b}\bar{m}^a\ell^b=i\M{2}{3}{0} $, $\gamma+\bar\gamma\equiv n_{a;b}\ell^an^b=-L_{11}$, $\gamma-\bar\gamma\equiv -\bar{m}_{a;b}m^an^b=i\M{2}{3}{1}$, $\sqrt{2}(\beta-\bar\alpha)\equiv \sqrt
{2}m_{a;b}\bar{m}^a{m}^b=\M{2}{3}{3}+i\M{2}{3}{2}$, $-\sqrt{2}(\beta+\bar\alpha)\equiv \sqrt{2}\ell_{a;b}n^a m^b=L_{12}-iL_{13}$.}
and which will be employed throughout the paper (see section~2.2 in \cite{Pravdaetal04} for details; rotation coefficients in a standard orthonormal frame are discussed, e.g., in \cite{Eisenhart49}).

We shall also employ covariant derivatives along the frame vectors
\BE
D \equiv \ell^a \nabla_a, \qquad \bigtriangleup  \equiv n^a \nabla_a, \qquad \delta_i \equiv m^{(i)a} \nabla_a .
\EE

\subsection{Transformation properties}

Lorentz transformations of the basis $\bm{a} $ can be described in terms of null rotations (with $\bl$ or $\bn$ fixed), spins (i.e., spatial rotations of the vectors $\bm{i} $) and boosts in the $\bl$-$\bn$ plane. Under {\em null rotations} with $\bl$ fixed
\BE
 \hbl=\bl, \qquad \hbn =\bn+z_i\bm{i} -\pul z^2\bl , \qquad \hbm{i} =\bm{i} -z_i\bl ,
    \label{nullrot}
\EE
where the $z_i$ are real functions and $z^2\equiv z_iz^i$, the rotation coefficients transform as 
\beqn
 & & \hspace{-2.3cm} \hL_{11}=L_{11}+z_i(L_{1i}+L_{i1})+z_iz_jL_{ij}-\pul z^2L_{10}-\pul z^2z_iL_{i0} , \qquad \hL_{10}=L_{10}+z_iL_{i0} , \nonumber \label{nullrot2} \\
 & & \hspace{-2.3cm} \hL_{1i}=L_{1i}-z_iL_{10}+z_jL_{ji}-z_iz_jL_{j0} , \qquad \hL_{i1}=L_{i1}+z_jL_{ij}-\pul z^2L_{i0} ,  \nonumber \\
 & & \hspace{-2.3cm} \hL_{i0}=L_{i0} , \qquad \hL_{ij}=L_{ij}-z_jL_{i0} , \nonumber \\
 & & \hspace{-2.3cm} \hN_{i1}=N_{i1}+z_jN_{ij}+z_iL_{11}+z_j\M{j}{i}{1}-\pul z^2(N_{i0}+L_{i1})+z_iz_j(L_{1j}+L_{j1}) \nonumber \\
 & & \hspace{-1cm} {}+z_jz_k\M{j}{i}{k}-\pul z^2(z_iL_{10}+z_jL_{ij}+z_j\M{j}{i}{0})+z_iz_jz_kL_{jk} \nonumber \\
 & & \hspace{-1cm} {}+\pul z^2\left(-z_iz_jL_{j0}+\pul z^2L_{i0}\right)+\T z_i+z_j\delta_jz_i-\pul z^2Dz_i , \\
 & & \hspace{-2.3cm} \hN_{i0}=N_{i0}+z_iL_{10}+z_j\M{j}{i}{0}+z_iz_jL_{j0}-\pul z^2L_{i0}+Dz_i , \nonumber \\
 & & \hspace{-2.3cm} \hN_{ij}=N_{ij}+z_iL_{1j}-z_jN_{i0}+z_k\M{k}{i}{j}-z_j(z_iL_{10}+z_k\M{k}{i}{0})+z_iz_kL_{kj}-\pul z^2L_{ij} \nonumber \\
 & & \hspace{-1cm} {}-z_iz_jz_kL_{k0} +\pul z^2z_jL_{i0}+\delta_jz_i-z_jDz_i , \nonumber \\
 & & \hspace{-2.3cm} \hM{i}{j}{1}=\M{i}{j}{1}+2z_{[j}L_{i]1}+z_k\M{i}{j}{k}+2z_kz_{[j}L_{i]k}-\pul z^2\M{i}{j}{0}-z^2z_{[j}L_{i]0} ,
  \nonumber \\
 & & \hspace{-2.3cm} \hM{i}{j}{0}=\M{i}{j}{0}+2z_{[j}L_{i]0} , \qquad \hM{i}{j}{k}=\M{i}{j}{k}+2z_{[j}L_{i]k}-z_k\M{i}{j}{0}+2z_kz_{[i}L_{j]0} , \nonumber 
\eeqn
whereas null rotations with $\bn$ fixed are obtained by interchanging $\bl\leftrightarrow\bn$,
$L\leftrightarrow N$ and $0\leftrightarrow 1$. 

Under {\em spins} 
\be
 \hbl =  \bl, \qquad \hbn = \bn, \qquad \hbm{i} =  X^{i}_{\ j} \bm{j} , 
 \label{spins}
\ee
where the $X^{i}_{\ j}$ are $(n-2)\times(n-2)$ orthogonal matrices, one gets
\beqn
 & & \hspace{-2.3cm} \hL_{11}=L_{11} , \quad \hL_{10}=L_{10} , \quad \hL_{1i}=X^{i}_{\ j}L_{1j} , \quad \hL_{i1}=X^{i}_{\ j}L_{j1} , \quad  \hL_{i0}=X^{i}_{\ j}L_{j0} , \nonumber \label{spins2} \\  
 & & \hspace{-2.3cm} \hL_{ij}=X^{i}_{\ k}X^{j}_{\ l}L_{kl} , \qquad \hN_{i1}=X^{i}_{\ j}N_{j1} , \quad \hN_{i0}=X^{i}_{\ j}N_{j0} , \quad   \hN_{ij}=X^{i}_{\ k}X^{j}_{\ l}N_{kl} , \nonumber \\
 & & \hspace{-2.3cm} \hM{i}{j}{1}=X^{i}_{\ k}X^{j}_{\ l}\M{k}{l}{1}+X^{j}_{\ k}\T X^{i}_{\ k} , \qquad \hM{i}{j}{0}=X^{i}_{\ k}X^{j}_{\ l}\M{k}{l}{0}+X^{j}_{\ k}D X^{i}_{\ k} , \\
 & & \hspace{-2.3cm} \hM{i}{j}{k}=X^{i}_{\ l}X^{j}_{\ m}X^{k}_{\ n}\M{l}{m}{n}+X^{j}_{\ m}X^{k}_{\ n}\delta_n X^{i}_{\ m}  . \nonumber 
\eeqn

Finally under {\em boosts} 
\be
 \hbl = \lambda \bl, \qquad  \hbn = \lambda^{-1} \bn, \qquad \hbm{i} = \bm{i} ,
 \label{boosts}
\ee
where $\lambda$ is a real function, one finds
\beqn
 & & \hspace{-2.3cm} \hL_{11}=\lambda^{-1}L_{11}+\lambda^{-2}\T\lambda , \quad \hL_{10}=\lambda L_{10}+D\lambda , \quad \hL_{1i}=L_{1i}+\lambda^{-1}\delta_i\lambda , \quad \hL_{i1}=L_{i1} , \nonumber \label{boosts2} \\ 
 & & \hspace{-2.3cm} \hL_{i0}=\lambda^2L_{i0} , \quad \hL_{ij}=\lambda L_{ij} , \qquad \hN_{i1}=\lambda^{-2}N_{i1} , \quad \hN_{i0}=N_{i0} , \quad   \hN_{ij}=\lambda^{-1}N_{ij} , \\
 & & \hspace{-2.3cm} \hM{i}{j}{1}=\lambda^{-1}\M{i}{j}{1} , \qquad \hM{i}{j}{0}=\lambda\M{i}{j}{0} ,\qquad \hM{i}{j}{k}=\M{i}{j}{k} . \nonumber 
\eeqn

\section{Ricci identities}

\label{sec_ricci}

Contractions of the Ricci identities $v_{a;bc}-v_{a;cb}={R}_{sabc}v^s$ with various  combinations of the frame vectors lead to the set of first order differential equations presented below, where ${\cal L}_{abc} $, ${\cal N}_{abc} $ and ${\cal M}_{abc} $ denote  the Ricci identities 
for $v^a=\ell^{a}$, $n^a$ and $m^{(i)a}$, respectively. 
For $n=4$ these are equivalent to the standard Ricci identities arising in the Newman-Penrose formalism \cite{Stephanibook}, and for $n=3$ to the equations of \cite{HalMorPer87}. 
 Note that some of these equations are related by interchanging $\bl\leftrightarrow\bn$,
$L\leftrightarrow N$, $D\leftrightarrow \bigtriangleup$  and $0\leftrightarrow 1$ and
in that case they are written  next to each other. 
\numparts
\bea
\fl 
{\cal L}_{abc} n^a \ell^b n^c \, : \   DL_{11} - \bigtriangleup L_{10} = -2 L_{10} L_{11} - L_{1i} L_{i1} 
+ N_{i0}(L_{1i}+ L_{i1}) - L_{i0} N_{i1} \nonumber \label{Lnln} \\
 \qquad\qquad\qquad {}- C_{0101} + \textstyle{\frac{2}{n-2}} R_{01} - \textstyle{\frac{1}{(n-1)(n-2)}}R\, ,  \\
%--------------------------------------------------------
\fl
{\cal L}_{abc} n^a \ell^b m^{(i) c}  \, : \   DL_{1i} - \delta_i L_{10} = - L_{11} L_{i0}  
- L_{10}(L_{1i}+ N_{i0}) + 2 L_{j[i|} N_{j|0]} \nonumber\\ 
 \qquad\qquad\qquad {}-L_{1j} (L_{ji}+\M{j}{i}{0}) - C_{010i} +  \textstyle{\frac{1}{n-2}}R_{0i}\, , \\
%--------------------------------------------------------
\fl
{\cal N}_{abc} \ell^a n^b m^{(i) c}  \, : \   \bigtriangleup L_{1i} - \delta_i L_{11} = - L_{10} N_{i1}  
+ L_{11}(L_{1i}- L_{i1}) - 2 L_{j[1|} N_{j|i]} \nonumber\\ 
 \qquad\qquad\qquad {}-L_{1j} (N_{ji}+\M{j}{i}{1}) + C_{101i} - \textstyle{\frac{1}{n-2}} R_{1i}\, , \\
%--------------------------------------------------------
\fl
{\cal L}_{abc} n^a m^{(i) b} m^{(j) c}  \, : \   \delta_{[j|} L_{1|i]} = - L_{11} L_{[ij]}  
- L_{10} N_{[ij]} -  L_{1k} \M{k}{[i}{j]} \nonumber\\ 
 \qquad\qquad\qquad {}- L_{k[j|} N_{k|i]} + \textstyle{\frac{1}{2}}C_{01ij}\, ,  \\
%--------------------------------------------------------
\fl
{\cal L}_{abc} m^{(i) a} \ell^{b} n^{c}  \, : \   D L_{i1} - \bigtriangleup L_{i0} = - 2L_{11} L_{i0}  
+ L_{ij} (-L_{j1} + N_{j0}) + 2 L_{j[0|} \M{j}{i|}{1]} \nonumber\\ 
 \qquad\qquad\qquad {}- C_{010i} + \textstyle{\frac{1}{n-2}}R_{0i}\, , \\
%--------------------------------------------------------
\fl
{\cal N}_{abc} m^{(i) a} n^{b} \ell^{c}  \, : \   \bigtriangleup N_{i0} - D N_{i1} =  2L_{10} N_{i1}  
+ N_{ij} (-N_{j0} + L_{j1}) + 2 N_{j[1|} \M{j}{i|}{0]} \nonumber\\ 
 \qquad\qquad\qquad {}- C_{101i} + \textstyle{\frac{1}{n-2}}R_{1i}\, , \\
%--------------------------------------------------------
\fl
{\cal L}_{abc} m^{(i) a} \ell^{b} m^{(j) c} \, : \   D L_{ij} - \delta_j L_{i0} = L_{10} L_{ij}  
- L_{i0} (2 L_{1j} + N_{j0}) - L_{i1} L_{j0}  
\nonumber\\
 \qquad\quad {}+ 2 L_{k[0|} \M{k}{i|}{j]} - L_{ik} (L_{kj} + \M{k}{j}{0})   
 - C_{0i0j} -\textstyle{\frac{1}{n-2}}R_{00} \delta_{ij}\, ,  \label{Lmlm}\\
 %--------------------------------------------------------
\fl
{\cal N}_{abc} m^{(i) a} n^{b} m^{(j) c} \, : \   \bigtriangleup N_{ij} - \delta_j N_{i1} = - L_{11} N_{ij}  
- N_{i1} (-2 L_{1j} + L_{j1}) - N_{i0} N_{j1}  
\nonumber\\
 \qquad\quad {}+ 2 N_{k[1|} \M{k}{i|}{j]} - N_{ik} (N_{kj} + \M{k}{j}{1})   
 - C_{1i1j} - \textstyle{\frac{1}{n-2}} R_{11}\delta_{ij}\, ,  \\
  %--------------------------------------------------------
\fl
{\cal L}_{abc} m^{(i) a} n^{b} m^{(j) c} \, : \   \bigtriangleup L_{ij} - \delta_j L_{i1} =  L_{11} L_{ij}  
- L_{i1} L_{j1} - L_{i0} N_{j1}   + 2 L_{k[1|} \M{k}{i|}{j]} 
\nonumber\\
  \quad \!\!\!\!\!\!\!\! \!\!\!\!\!\!\! {}- L_{ik} (N_{kj} + \M{k}{j}{1})   
 - C_{0i1j} -  \textstyle{\frac{1}{n-2}} (R_{ij}+R_{01} \delta_{ij})  + \textstyle{\frac{1}{(n-1)(n-2)}} R \delta_{ij} \, ,  \\
  %--------------------------------------------------------
\fl
{\cal N}_{abc} m^{(i) a} \ell^{b} m^{(j) c} \, : \   D N_{ij} - \delta_j N_{i0} = - L_{10} N_{ij}  
- N_{i0} N_{j0} 
 %\nonumber\\
 - N_{i1}  L_{j0}   + 2 N_{k[0|} \M{k}{i|}{j]}  \nonumber\\
  \quad \!\!\!\!\!\!\!\! \!\!\!\!\!\!\!  {}- N_{ik} (L_{kj} + \M{k}{j}{0})   %\nonumber\\
 - C_{0j1i} - {\textstyle{\frac{1}{n-2}}} (R_{ij}+R_{01}\delta_{ij})+  \textstyle{\frac{1}{(n-1)(n-2)}} R \delta_{ij} \, , \\
  %--------------------------------------------------------
\fl
{\cal L}_{abc} m^{(i) a} m^{(j)b} m^{(k) c} \, : \    \delta_{[j|} L_{i|k]} =  L_{1[j|} L_{i|k]}  
+ L_{i1} L_{[jk]} + L_{i0} N_{[jk]}   +  L_{il} \M{l}{[j}{k]} 
\nonumber\\
  \qquad\qquad\qquad\qquad \!\!\!\!\!\!\!\! \!\!\!\!\!\!\ {}+ L_{l[j|}  \M{l}{i|}{k]}   
 - \textstyle{\frac{1}{2}} C_{0ijk} - \textstyle{\frac{1}{n-2}} R_{0[j} \delta_{k]i}\, , \label{Lmmm}  \\
  %-------------------------------------------------------- 
\fl
{\cal N}_{abc} m^{(i) a} m^{(j)b} m^{(k) c} \, : \    \delta_{[j|} N_{i|k]} =  - L_{1[j|} N_{i|k]}  
+ N_{i0} N_{[jk]} + N_{i1} L_{[jk]}   +  N_{il} \M{l}{[j}{k]} 
\nonumber\\
  \qquad\qquad\qquad\qquad \!\!\!\!\!\!\!\! \!\!\!\!\!\!\ {}+ N_{l[j|}  \M{l}{i|}{k]}   
 - \textstyle{\frac{1}{2}}C_{1ijk} - \textstyle{\frac{1}{n-2}} R_{1[j} \delta_{k]i}\, , \\ 
 %--------------------------------------------------------
\fl
{\cal M}_{abc} m^{(j)a} \ell^b n^{ c}  \, : \   D \M{i}{j}{1} - \bigtriangleup \M{i}{j}{0} =  2N_{i[1|} L_{j|0]}  
+2 L_{i[1|} N_{j|0]} -L_{10} \M{i}{j}{1} 
 \nonumber\\ 
 \qquad\quad \!\!\! {}-L_{11} \M{i}{j}{0}
 + 2  \M{i}{k}{[0|} \M{k}{j|}{1]} + \M{i}{j}{k} (-L_{k1} +N_{k0}) 
 - C_{01ij}\, ,  \\
  %--------------------------------------------------------
\fl
{\cal M}_{abc} m^{(j)a} \ell^b m^{(k) c}  \, : \   D \M{i}{j}{k} - \delta_{k} \M{i}{j}{0} 
=  2L_{j[0|} N_{i|k]}  
+2 N_{j[0|} L_{i|k]} -L_{k0} \M{i}{j}{1} 
 \nonumber\\ 
  \! \! \! \! \! \! \! \! \! \! \! \!  \! \! \! \! \! \!  \! \! \! \! \! \! 
\! \! \! \! \! \! \!
\qquad\qquad\qquad\qquad\quad {}- \M{i}{j}{0} (L_{1k}+N_{k0})
+ 2  \M{i}{l}{[0|} \M{l}{j|}{k]} - \M{i}{j}{l} (L_{lk} + \M{l}{k}{0}  )  \nonumber\\ 
  \! \! \! \! \! \! \! \! \! \! \! \!  \! \! \! \! \! \!  \! \! \! \! \! \! 
\! \! \! \! \! \! \!
\qquad\qquad\qquad\qquad\quad {}- C_{0kij} -\textstyle{\frac{1}{n-2}}{2R_{0[i}\delta_{j]k}}\, ,   \\
   %--------------------------------------------------------
\fl
{\cal M}_{abc} m^{(j)a} n^b m^{(k) c}  \, : \   \bigtriangleup \M{i}{j}{k} - \delta_{k} \M{i}{j}{1} 
=  2N_{j[1|} L_{i|k]}  
+2 L_{j[1|} N_{i|k]} -N_{k1} \M{i}{j}{0} 
 \nonumber\\ 
 \! \! \! \! \! \! \! \! \! \! \! \!  \! \! \! \! \! \!  \! \! \! \! \! \! 
\! \! \! \! \! \! \!
\qquad\qquad\qquad\qquad\quad
 {}+ \M{i}{j}{1} (L_{1k}-L_{k1})
+ 2  \M{i}{l}{[1|} \M{l}{j|}{k]} - \M{i}{j}{l} (N_{lk} + \M{l}{k}{1}  )  \nonumber\\ 
 \! \! \! \! \! \! \! \! \! \! \! \!  \! \! \! \! \! \!  \! \! \! \! \! \! 
\! \! \! \! \! \! \!
\qquad\qquad\qquad\qquad\quad {}- C_{1kij} -\textstyle{\frac{1}{n-2}}{ 2R_{1[i}\delta_{j]k} }\, ,   \\
    %--------------------------------------------------------
\fl
{\cal M}_{abc} m^{(j)a} m^{(k)b} m^{(l) c}  \, : \   \delta_{[k|} \M{i}{j|}{l]} 
=  N_{i[l|} L_{j|k]}  
+ L_{i[l|} N_{j|k]} +L_{[kl]} \M{i}{j}{1} +N_{[kl]} \M{i}{j}{0}\nonumber \label{Mmmm}\\ 
{}+   \M{i}{p}{[k|} \M{p}{j|}{l]} + \M{i}{j}{p}  \M{p}{[k}{l]} 
  - \textstyle{\frac{1}{2}} C_{ijkl} -\textstyle{\frac{1}{n-2}} (\delta_{i[k}  R_{l]j}-\delta_{j[k}R_{l]i}) \nonumber \\
  {}+\textstyle{\frac{1}{(n-1)(n-2)}}R\delta_{i[k}\delta_{l]j} \, .
\eea
\endnumparts

Further contractions do not lead to new equations because of eqs.~(\ref{products}) (for example, taking the second derivative of $\ell_a n^a =1$ leads to $n_{a;[bc]}\ell^a + \ell_{a;[bc]}n^a=0$, so that ${\cal L}_{abc} n^a$ and ${\cal N}_{abc} \ell^a$ are identical; other normalization conditions give similar constraints). 

\section{Null geodetic congruences and Sachs equations}

\label{sec_sachs}

In this section we consider the physically interesting case of a geodetic vector $\bl$. By~(\ref{derlnm}), $\ell_{a ;b}\ell^b=L_{10}\ell_a+L_{i0}m^{(i)}_a$, so that $\bl$ is geodetic if and only if $L_{i0}=0$. In this case, the matrix $L_{ij}$ acquires a special meaning since it is then invariant under the null rotations~(\ref{nullrot}) preserving $\bl$, see eq.~(\ref{nullrot2}) (and it simply rescales with boost weight one under a boost~(\ref{boosts}), cf.~eq.~(\ref{boosts2})). It is then convenient to
decompose $L_{ij}$ into its tracefree symmetric part $\sigma_{ij}$ ({\em shear}), its trace $\theta$ ({\em expansion}) and its antisymmetric part $A_{ij}$ ({\em twist}) as \cite{Pravdaetal04} (cf. also \cite{FroSto03,LewPaw05}) 
\beqn
 & & L_{ij}=\sigma_{ij}+\theta\delta_{ij}+A_{ij} , \label{L_decomp} \\
 & & \sigma_{ij}\equiv L_{(ij)}-\textstyle{\frac{1}{n-2}}L_{kk}\delta_{ij} , 
\qquad \theta\equiv\textstyle{\frac{1}{n-2}}{L_{kk}} , \qquad A_{ij}\equiv L_{[ij]} . \label{opt_matrices}
\eeqn 

Along with $\theta$, one can construct other scalar quantities out of $\ell_{a ; b}$ (see also the footnote 
on p.~\pageref{note_invariants}) 
which are invariant under null and spatial rotations with fixed $\bl$, e.g. the shear and twist scalars given by the traces $\sigma^2\equiv\sigma^2_{ii}=\sigma_{ij}\sigma_{ji}$ and $\omega^2\equiv-A^2_{ii}=-A_{ij}A_{ji}$ (note that $\sigma_{ij}=0\Leftrightarrow\sigma^2=0$ and
$A_{ij}=0\Leftrightarrow\omega^2=0$). 

If $\bl$ is {\em affinely parametrized}, i.e. $L_{10}=0$ (as we shall assume in the following), the optical scalars take the form
\be
 \hspace{-1.3cm} \sigma^2=\ell_{(a;b)}\ell^{(a;b)}-\textstyle{\frac{1}{n-2}}\left(\ell^a_{\;;a}\right)^2 , \qquad \theta=\textstyle{\frac{1}{n-2}}\ell^a_{\;;a} , \qquad 
 \omega^2=\ell_{[a;b]}\ell^{a;b} .
\ee

It is worth observing that one can always choose $\bn$ and $\bm{i} $ to be {\em parallely propagated} along the geodesic null congruence $\bl$. Namely, one can set $\M{i}{j}{0}=0$ and $N_{i0}=0$ by performing an appropriate spin transformation~(\ref{spins}) and then a null rotation~(\ref{nullrot}). 
This simplifies  the Ricci identities considerably and may be convenient for certain calculations, but it will not be assumed in the rest of this paper.

\subsection{Sachs equations}

Now, setting $L_{i0}=0=L_{10}$ in~(\ref{Lmlm}) and splitting this identity into its tracefree symmetric part, its trace and its antisymmetric part according to eq.~(\ref{L_decomp}), one ends up with the set of $n$-dimensional Sachs equations 
\numparts
\beqn
 & & \fl D\sigma_{ij}=-\left({\sigma^2}_{ij}-\textstyle{\frac{1}{n-2}}\sigma^2\delta_{ij}\right)-\left({A}^2_{ij}+\textstyle{\frac{1}{n-2}}\omega^2\delta_{ij}\right)-2\theta\sigma_{ij}-2\sigma_{k(i}\M{k}{j)}{0}-C_{0i0j} , \label{sachs_she} \\ 
 & & \fl D\theta=-\textstyle{\frac{1}{n-2}}\sigma^2-\theta^2+\textstyle{\frac{1}{n-2}}\omega^2-\textstyle{\frac{1}{n-2}}R_{00} , \label{sachs_exp} \\
 & & \fl DA_{ij}=-2\theta A_{ij}-2\sigma_{k[j}A_{i]k}+2A_{k[i}\M{k}{j]}{0} . \label{sachs_twi}
\eeqn
\endnumparts

Notice that for $n=4$ one has identically ${\sigma^2}_{ij}-\textstyle{\frac{1}{n-2}}\sigma^2\delta_{ij}=0={A}^2_{ij}+\frac{1}{n-2}\omega^2\delta_{ij}$ and $\sigma_{k[j}A_{i]k}=0=2A_{k[i}\M{k}{j]}{0}$ so that eqs.~(\ref{sachs_she}) and (\ref{sachs_twi}) simplify considerably and take the standard form \cite{Sachs61,Stephanibook,penrosebook2}. Eq.~(\ref{sachs_exp}) is instead essentially unchanged, and for $n\ge 4$ has been presented previously in \cite{LewPaw05,PodOrt06}.

From eqs.~(\ref{sachs_she}) and (\ref{sachs_twi}) one can also obtain the following scalar equations (which characterize the propagation of $\bl$ indepentently of the other frame vectors\footnote{\label{note_invariants} Note indeed that 
 $\sigma^3\equiv\sigma_{ij}\sigma_{jk}\sigma_{ki}=g^{ad}\ell_{(a;b)}\ell_{(c;d)}\ell^{(b;c)}-3\theta\sigma^2-(n-2)\theta^3$, 
$\sigma_{ij}A^2_{ji}=\sigma_{ij}A_{jk}A_{ki}=g^{ad}\ell_{(a;b)}\ell_{[c;d]}\ell^{[b;c]}+\theta A^2$ and $C_{0i0j}\sigma_{ij}=C_{abcd}\ell^a\ell^c\ell^{b;d}$. The higher order optical scalars $\sigma^3$ and $\sigma_{ij}A^2_{ji}$ entering (\ref{sachs_she2}) and (\ref{sachs_twi2}) vanish in $n=4$. However, for $n>4$ these and other higher order optical scalars may be in general non-zero and independent of $\theta$, $\sigma$ and $\omega$.})
\numparts
\beqn
 & & D(\sigma^2)=-2\sigma^3-2\sigma_{ij}A^2_{ji}-4\theta\sigma^2-2C_{0i0j}\sigma_{ij} \label{sachs_she2} , \\
 & & D(\omega^2)=-4\theta \omega^2+4\sigma_{ij}A^2_{ji} . \label{sachs_twi2}
\eeqn 
\endnumparts 

In the special case $\sigma=0=R_{00}$ one can integrate (\ref{sachs_exp}) and (\ref{sachs_twi2}) as in \cite{NP,penrosebook2} to obtain 
\be
\fl  \theta=\frac{\theta_0+r\left(\theta_0^2+\textstyle{\frac{1}{n-2}}\omega_0^2\right)}{1+2r\theta_0+r^2\left(\theta_0^2+\textstyle{\frac{1}{n-2}}\omega_0^2\right)} , \qquad \omega=\frac{\omega_0}{1+2r\theta_0+r^2\left(\theta_0^2+\textstyle{\frac{1}{n-2}}\omega_0^2\right)} ,
\ee
where $r$ is an affine parameter along $\bl$ and $\theta_0$, $\omega_0$ are the values of the scalars at $r=0$ (in the non-twisting case, $\omega_0=0$, this reduces to the result of~\cite{PodOrt06}). 

\subsection{Consequences of the Sachs equations}

A number of simple but important facts readily follow from the above equations. First, as in $n=4$, it is the Ricci component $R_{00}$  of boost weight 2 which controls the propagation of expansion/convergence, and the Weyl components $C_{0i0j}$ (also of boost weight 2) which control the propagation of shear. On the other hand, the curvature tensor does not enter the twist-propagation equation~(\ref{sachs_twi}).

From eqs.~(\ref{sachs_she}) and (\ref{sachs_exp}) one immediately finds
\begin{proposition}
\label{prop_wand}
 Given a geodetic null congruence with tangent vector $\bl$ in an arbitrary $n$-dimensional spacetime ($n\ge 4$), the following implications hold:
\begin{enumerate}
 \item $R_{00}=0$, $\theta=0=\sigma_{ij}$ $\ \Rightarrow \ $ $A_{ij}=0$, $C_{0i0j}=0$. \label{expshe}
 \item $R_{00}=0$, $\theta=0=A_{ij}$ $\ \Rightarrow \ $ $\sigma_{ij}=0$, $C_{0i0j}=0$. \label{exptwi}
\end{enumerate}
 That is, when $R_{00}=0$ if a geodetic null congruence is non-expanding and shearfree it must be also twistfree, and it is  automatically a WAND; if it is non-expanding and twistfree it must be also shearfree, and again a WAND.
\end{proposition}

These are well-known properties in $n=4$ \cite{penrosebook2}. For $n>4$, the first conclusion in~(\ref{exptwi}) was already pointed out in \cite{LewPaw05}. 
Note that the assumed condition $R_{00}=0$ on the matter content is satisfied in a large class of spacetimes such as vacuum (with a possible cosmological constant), aligned pure radiation and aligned Maxwell fields. 

It is worth observing that both the alternative assumptions in~(\ref{expshe}) and (\ref{exptwi}) uniquely identify the Kundt class of non-expanding, twistfree and shearfree spacetimes, i.e. with $L_{ij}=0$ (see \cite{ColHerPel06,Coleyetal06} and references therein for related studies in higher dimensions). In view of~(\ref{exptwi}), we conclude that one can not ``generalize'' the Kundt solutions by allowing for a non-zero shear (as long as $R_{00}=0$ and one insists on the non-expanding and twistfree conditions). This should be contrasted with the existence of ``generalized Robinson-Trautman solutions'' with non-zero shear in $n>4$ (which can be readily obtained by taking the direct product of a four-dimensional Robinson-Trautman solution with a flat $\mathbb{R}^q$, $q\ge 1$; see instead \cite{PodOrt06} for standard Robinson-Trautman $n>4$ solutions). Furthermore, in both  above cases (\ref{expshe}) and (\ref{exptwi}), the fact that the tangent vector $\bl$ is necessarily a WAND (because of $C_{0i0j}=0$) implies for $n>4$ that the considered spacetime is algebraically special, i.e. it can not be of type G (this was already noticed in \cite{Pravdaetal04}). In addition, if we now substitute $L_{i0}=0=L_{ij}$ in (\ref{Lmmm}) and we further assume $R_{0i}=0$, we obtain $C_{0ijk}=0$. Recalling the identity $C_{010i}=C_{0jij}$ \cite{Pravdaetal04} we find also $C_{010i}=0$, so that with Proposition~\ref{prop_wand} we conclude:
\begin{proposition}
\label{prop_kundt}
 Under the assumption $R_{00}=0=R_{0i}$ on the matter fields, $n\ge 4$ dimensional Kundt spacetimes (i.e., $L_{i0}=0=L_{ij}$) are of type II (or more special).
\end{proposition} 

Next, one can observe that for odd $n$, ${A}^2_{ij}+\frac{1}{n-2}\omega^2\delta_{ij}=0$ $\Leftrightarrow$ $A_{ij}=0$ (since one needs $(-\frac{1}{n-2}\omega^2)^{n-2}=\det({A}^2_{ij})$, and $\det({A}^2_{ij})=0$ for odd $n$). Thus, it follows from eq.~(\ref{sachs_she}) that
\begin{proposition}

In an  $n>4$  dimensional spacetime, {$n$ odd}, for a geodetic WAND \footnote{Recall that in $n=4$, by the Goldberg-Sachs theorem \cite{GolSac62,NP,Stephanibook,penrosebook2}, a multiple principal null direction of a vacuum gravitational fields is necessarily geodetic. Similarly, for $n>4$ multiple WANDs are necessarily geodetic in the case of type III and type N spacetimes \cite{Pravdaetal04}. In \cite{PraPraOrt07} we shall present a partial extension of this property to type II and D vacuum spacetimes (under further assumptions on the Weyl tensor). However, we shall also point out that a complete extension is not possible, namely there do exist multiple WANDs that are non-geodetic (as opposed to the case $n=4$).}
 \[
 A_{ij}\neq 0 \ \Rightarrow \  \sigma_{ij}\neq 0 ,
\]
i.e., a twisting geodetic WAND must also be shearing. Conversely, a twisting non-shearing geodetic null direction 
cannot be a WAND.
\end{proposition}

The above result is a counterexample to a complete higher dimensional extension of the Goldberg-Sachs theorem, according to which algebraically special vacuum spacetimes are shearfree in $n=4$. Myers-Perry black holes in $n=5$ provide an explicit example of algebraically special (type D) twisting and  shearing spacetimes \cite{FroSto03,Pravdaetal04}. Note that the converse implication $\sigma_{ij}\neq 0 \ \Rightarrow \  A_{ij}\neq 0$ does not hold: for example, static black strings/branes (i.e., the product ($p$-dimensional Schwarzschild)$\times\mathbb{R}^q$, with $p\ge 4$ and $q\ge 1$) are type D spacetimes with  expanding, shearing but non-twisting  geodetic multiple  WANDs \cite{PraPraOrt07}.

On the other hand, again using eq.~(\ref{sachs_she}) one has
\begin{proposition}

\label{lem_}

In any spacetime with $n>4$, given a geodetic null congruence which is {\em not} a WAND 
(in particular, an arbitrary geodetic null congruence in an algebraically general (type G) spacetime)
\[
 \sigma_{ij}=0 \ \Rightarrow \  A_{ij}\neq 0 , 
\] 
i.e., a shearfree null geodesic is necessarily twisting unless it is a WAND.
\end{proposition}

\ack
V.P. and A.P. acknowledge support from research plan No
AV0Z10190503 and research grant KJB100190702.

\section*{References}

%\bibliographystyle{unsrt}

%\bibliography{bibl,Vojta}

\end{document}